\documentstyle[aps,epsfig,prl,floats]{revtex}

\newcommand{\be}{\begin{equation}}
\newcommand{\ee}{\end{equation}}
\newcommand{\bea}{\begin{eqnarray}}
\newcommand{\eea}{\end{eqnarray}}

\tightenlines
\begin{document}
\draft

\title{Linked Cluster Series Expansions for Two-Particle Bound States} 
\author{Weihong Zheng\cite{zwh} and Chris J.~Hamer \cite{cjh}}
\address{
School of Physics, University of New South Wales, Sydney NSW 2052, Australia}
\author{Rajiv R.~P.~Singh}
\address{
Department of Physics, University of California, Davis, CA 95616}
\author{Simon Trebst \cite{Bell} and Hartmut Monien}
\address{
Physikalisches Institut, Universit\"at Bonn, Nu\ss allee 12, 53115 Bonn, 
Germany}
           
\date{\today}

\maketitle

\begin{abstract}

We develop strong-coupling series expansion methods to study two-particle spectra
of quantum lattice models. At the heart of the method lies the calculation of 
an effective Hamiltonian in the two-particle subspace. We explicitly consider
an orthogonality transformation to generate this block diagonalization, and 
find that maintaining orthogonality is crucial for systems 
where the ground state and the
two-particle subspace are characterized by identical quantum numbers. 
We discuss the solution of the two-particle Schr\"odinger equation 
by using a finite lattice approach in coordinate space 
or by an integral equation in momentum space.
These methods allow us to 
precisely determine the low-lying excitation spectra of the models at hand,
including all two-particle bound/antibound states.
Further, we discuss how to generate series expansions for the dispersions of 
the bound/antibound states. These allow us to employ series 
extrapolation techniques, whereby binding energies can be determined
even when the expansion parameters are not small.
We apply the method to the (1+1)D transverse Ising model and the two-leg 
spin-$\case 1/2$ Heisenberg ladder. For the latter model, we also
calculate the coherence lengths and determine the critical properties
where bound states merge with the two-particle continuum.

\end{abstract}


\section{INTRODUCTION}

The study of bound states and multiparticle excitations remains a challenging
problem in many-body physics. Experimentally, there are several probes for
low-dimensional magnetic or strongly   correlated electronic systems which
show spectral features associated with multiparticle continuum and
bound states. These include two-magnon Raman spectra, optical absorption,
photoemission and neutron scattering spectra. The multiparticle
features often remain poorly understood. On the theoretical side, one example
of the intriguing issues that may arise is the role that the increasing
number of bound states play in the confinement-deconfinement transition in spin-Peierls
systems. At the transition the spectrum switches from a soliton-antisoliton continuum to
elementary triplet excitations, their bound states and continuum\cite{Affleck:97}.

A controlled numerical framework for the calculation of multi-particle
spectral properties, which can also account for various singularities as the
coupling constants are varied, is currently missing.  In one dimension,
a variety of numerical methods including Lanczos, exact
diagonalization and most notably density matrix renormalization group
(DMRG) \cite{KuehnerWhite99} hold promise for such
calculations. However, unlike ground state and single-particle
properties, the calculation of full dynamical properties like spectral
functions still needs more conceptual advances.  In higher than one
dimension, all of these methods are restricted to small system sizes,
which makes it difficult to study the thermodynamic limit.

On the other hand in the limits of weak or strong couplings,
perturbation theory can be used to calculate all properties of the
multiparticle spectra directly in the thermodynamic limit. If these
calculations can be done to high orders, one can calculate
multiparticle spectra in a systematic manner using extrapolation
techniques even in cases where the perturbations are not weak. In
principle, one can see that as the couplings are increased the number
of bound states can change and states can come off or merge into the
continuum.  The resulting singularities should be amenable to series
expansion methods.

In this paper, we show how to calculate multi-particle spectral properties from
high-order perturbation expansions, using a linked cluster method.
A brief outline and summary of the work was given in a recent letter\cite{shortpaper}.
Our method is quite distinct from the flow equation approach of Wegner\cite{flow}, 
which has also been used recently by Uhrig and collaborators \cite{Uhrig:98,Knetter:00} 
for the study of multiparticle spectral properties in one and two dimensions.

	The linked cluster method is by far the most efficient way to
generate perturbation series expansions for quantum Hamiltonian lattice
models. For the ground state energy and related properties, a
linked cluster approach was first discussed in unpublished work by
Nickel\cite{nickel}, followed by work of Marland\cite{marland}, Irving
and Hamer\cite{irving1} and others, as reviewed by He et al\cite{he}.
The approach was later rediscovered and applied to a whole new range of
problems in condensed matter physics by Singh, Huse and
Gelfand\cite{singh1,gelfand1}.

	For the energies of excited states, it is more difficult to
formulate a true linked cluster expansion, although related methods have
been known for some time\cite{nickel,hamer}. It was only in 1996
that the key to a true linked cluster expansion for one-particle excited
states was discovered by Gelfand\cite{gelfand2}. Since then, many
applications of this technique have been made, calculating
single-particle energies, dispersion relations and spectral functions
in models of interest in condensed matter physics. 
For a recent review, see Gelfand and Singh\cite{gelfand4}.

The key advantage of the cluster expansion
method is that the calculations can be carried out systematically and
efficiently by fully automated computer programs.  Furthermore, these
methods work by breaking up the thermodynamic problem into a purely
combinatorial problem and a number of finite-cluster problems. Thus,
while they are technically harder in higher than one dimension, the
difficulty is not fundamental. In fact, over the years, a number of
workers\cite{jaanwork,bonnsite,gelfand4,martin,rap74} have
independently developed efficient computer programs
to  generate these clusters automatically,
and the cluster data up to quite high number of vertices
for most 2-dimensional and 3-dimensional lattices 
including the simple-cubic, BCC and FCC lattices have been 
generated\cite{jaanwork}: these data
can be applied to a wide range of models.

At the heart of our new approach is a generalization of Gelfand's 
linked cluster expansion for single-particle excited states to two-particle
states. From a technical point of view, our most notable achievement is the
development of an orthogonality transformation which leads to a linked cluster
theorem for multi-particle states even when their quantum numbers are identical
with the ground state. We show how to calculate energies and dispersion
relations for two-particle excitations, and coherence lengths for the bound
states. The further generalization to higher number of particle is then
obvious in principle.

As a first check to ensure that the method is working correctly, we apply it to the
case of the transverse Ising model in one dimension, which can be solved exactly
in term of free fermions\cite{exactIsing}. We show that the series for the
2-particle state agree with the exact results up to 12th order.

Finally, we apply the method to a non-trivial model, the 2-leg spin-$\case 1/2$
Heisenberg ladder, which has been much discussed 
recently\cite{Dag96,Gopalan:94,Oitmaa:96,Eder:97,Damle:98,Jurecka:99,Kotov:98,Kotov:99}
as a prime example of a one-dimensional antiferromagnetic system with a gapped
excitation spectrum. The two-particle bound states have already been studied by Damle and Sachdev\cite{Damle:98}
and Sushkov and Kotov\cite{Kotov:98,Kotov:99}. We perform a detailed study of these bound states,
exhibiting in particular the characteristic features as each bound state emerges
from the continuum.

In a companion paper\cite{longj12d}, we apply the same techniques to a still
more interesting case, the frustrated alternating
Heisenberg chain, which displays the 
confinement-deconfinement transition discussed by Affleck\cite{Affleck:97}.

The organization of the paper is as follows.
Section II of the paper lays out the formalism and methods used to obtain
the 2-particle spectra. Section III discusses the applications to the transverse
Ising chain and the Heisenberg ladder, and Section IV summarizes our conclusions.

\section{FORMALISM}

	We consider a Hamiltonian
\begin{equation}
               H = H_{0} + \lambda H_{1} \;,
\end{equation}
where the unperturbed Hamiltonian $H_{0}$ is exactly solvable, and
$\lambda$ is the perturbation parameter. 
In the lattice models of interest here, $H_{0}$ will typically consist
of single-site operators, while interaction terms between different
sites will be included in the perturbation operator $H_{1}$.
The aim is to calculate
perturbation series in $\lambda$ for the eigenvalues of $H$ and other
quantities of interest. 
The calculation proceeds in three stages.

\subsection{Block Diagonalization}
\label{secIIA}

	On any finite lattice or cluster of sites, the first step is to 
`block diagonalize' the Hamiltonian 
to form an ``effective Hamiltonian", where the ground state sits in a
block by itself, the 1-particle states form another block, the
2-particle states another block, and so on. 
Here a ``particle" may refer to a lattice fermion, a spin-flip, or other
excitation, depending on the model at hand.
We assume that all the unperturbed states in each block are degenerate
under $H_{0}$.
There is no unique way to block diagonalize the Hamiltonian,
but the eigenvalues and final results should be unique, independent 
of the method used,
 as long as the cluster expansion works correctly.
Gelfand\cite{gelfand2}
used a similarity transform for this purpose:
\begin{equation}
H^{\rm eff} = O^{-1} H O \;.
\end{equation}
This works correctly for most 1-particle problems, 
and also for those 2-particle  states which have different 
quantum numbers to the ground state. However in general, 
especially for the 2-particle states which have identical 
quantum numbers to the ground state,
we need to be a little more careful than this, in order to
preserve all the proper symmetries of the Hamiltonian. We must
ensure that the transformation is {\it unitary}. Here we will only consider
the case when the Hamiltonian is real symmetric, and can be block
diagonalized by an {\it orthogonal} transformation,
\begin{equation}
H^{\rm eff} = O^{T} H O
\end{equation}
or more conveniently
\begin{equation}
OH^{\rm eff} =  H O \;,
\end{equation}
where
\begin{equation}
O^{T} = O^{-1} \;.
\label{eq1.8}
\end{equation}
The orthogonality of O can be ensured by writing
\begin{equation}
O = e^{S} \;,
\end{equation}
where S is real, antisymmetric
\begin{equation}
S^{T} = -S \;.
\label{1.9} \end{equation}
This transformation is constructed order-by-order in perturbation
theory.
The matrix elements of $H^{\rm eff}$ between different blocks are
zero, up to the given order in perturbation theory. 
Each matrix is expanded in powers of $\lambda$:
\begin{equation}
O = \sum_{n=0}^{\infty} \lambda^{n}O^{n)} \;,
\end{equation}
\begin{equation}
S = \sum_{n=0}^{\infty} \lambda^{n}S^{n)} \;,
\end{equation}
\begin{equation}
H^{\rm eff} = \sum_{n=0}^{\infty} \lambda^{n}H_{\rm eff}^{n)} \;,
\end{equation}
where at zeroth order we set
\begin{equation}
S^{0)} = 0, \hspace{5mm}O^{0)} = I, \hspace{5mm}H_{\rm eff}^{0)} = H_{0} \;.
\end{equation}
where $I$ is an unit matrix, $H_{0}$ is a diagonal matrix, with diagonal matrix elements $E_{i}^{0}$. 

At higher orders $n \neq 0$, we have
\begin{equation}
O^{n)} = S^{n)} + \frac{1}{2}\sum_{m,l=1}^{n} S^{m)}S^{l)}\delta_{m+l,n}
+ \frac{1}{3!}\sum_{m,l,k=1}^{n}S^{m)}S^{l)}S^{k)}\delta_{m+l+k,n} + ...
\label{eq1.14}
\end{equation}
and
\begin{equation}
\sum^{n}_{m,l =0}O^{m)}H_{\rm eff}^{l)}\delta_{m+l,n} = H_{0}O^{n)} + H_{1}O^{n-1)}
\label{1.15}
 \end{equation}
and it is convenient to define
\begin{equation}
R^{n)} = O^{n)} -S^{n)} \;.
 \end{equation}

\begin{figure}
 \begin{center} 
 \vskip -1cm
 \epsfig{file=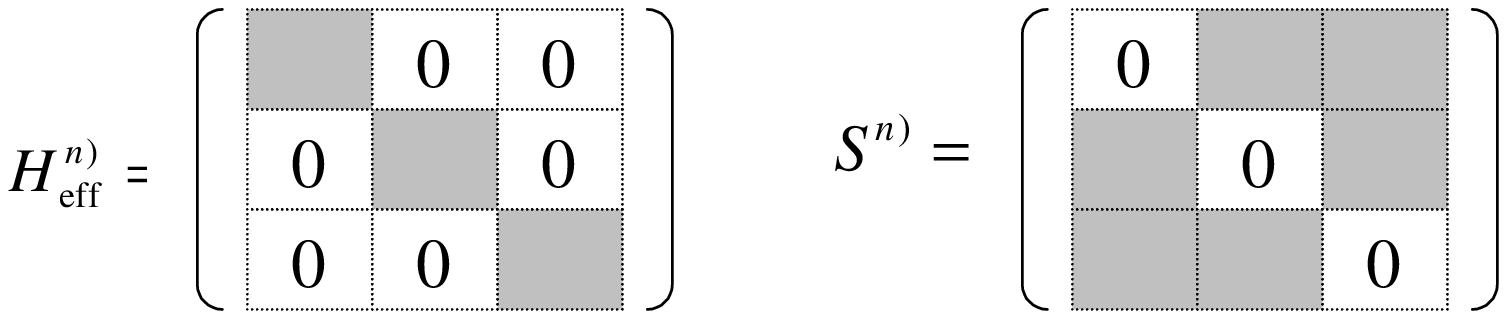, height=20cm}
  \vskip -17cm
 \caption[]
         {Block structure of the matrices $H^{n)}_{\rm eff}$ and $S^{n)}$. Setting the
upper right blocks of $H^{n)}_{\rm eff}$ to zero determines the
corresponding (shaded) blocks of $S^{n)}$; the diagonal blocks of $S^{n)}$ 
are set
to zero.}
 \label{fig1} 
 \end{center}   
\end{figure}

If we demand that at any given order n the off-diagonal blocks of
$H^{\rm eff}$ in (say) the upper right triangle vanish, then
equations (\ref{1.15}) determine the entries in the corresponding
blocks of $S^{n)}$ (Figure \ref{fig1}). The transposed blocks in the lower left
triangle are then determined by the antisymmetry condition (\ref{1.9}); and
only the diagonal blocks of $S$ remain to be determined. The simplest
choice is to set the diagonal blocks to zero.
Thus $S^{n)}$ is
completely determined:
\begin{eqnarray}
S^{n)}_{ij} = -R^{n)}_{ij} + \frac{1}{(E^{0}_{j}-E^{0}_{i})} \{ 
H_{1}O^{n-1)} -
  \sum_{m,l=1}^{n-1}
O^{m)}H^{l)}_{\rm eff}\delta_{m+l,n} 
\}_{ij}
\label{eq1.16}
\end{eqnarray}
or
\begin{eqnarray}
O^{n)}_{ij} =  \frac{1}{(E^{0}_{j}-E^{0}_{i})} \{ 
H_{1}O^{n-1)} -
  \sum_{m,l=1}^{n-1}
O^{m)}H^{l)}_{\rm eff}\delta_{m+l,n} 
\}_{ij}
\label{eq1.17}
\end{eqnarray}
for elements $ij$ in the off-diagonal (shaded) blocks. Then
\begin{eqnarray}
(H^{n)}_{\rm eff})_{ij} = 
\{H_{1}O^{n-1)} - 
  \sum_{m,l=1}^{n-1}O^{m)}H^{l)}_{\rm eff}\delta_{m+l,n} 
\}_{ij}
\label{eq1.18}
\end{eqnarray}
for elements in the diagonal blocks.
The right-hand sides of equations (\ref{eq1.16},\ref{eq1.17},\ref{eq1.18}) can 
all be computed
from the results at order $(n-1)$.

The key differences here from the similarity 
transformation are as follows. In the similarity transformation, the diagonal blocks
of $O^{n)}$ are undetermined, and so are chosen to be zero, while the off-diagonal
blocks of $O^{n)}$ are antisymmetric and can be determined by demanding 
the off-diagonal blocks of $H^{n)}_{\rm eff}$ to be zero. In the orthogonal
transformation, on the other hand, the diagonal blocks of $O^{n)}$ cannot
be chosen to be zero.
Instead the diagonal blocks of $S^{n)}$ are chosen to be zero, while 
the diagonal blocks of $O^{n)}$ are required to be nonzero by
orthogonality, and are determined by Eq. (\ref{eq1.14}).

At the end of this process, the effective Hamiltonian has been block
diagonalized, up to a given order in perturbation theory. The
orthogonal transformation will transform the unperturbed two-particle
states into ``dressed" states containing admixtures of different
particle numbers; and in particular, there will be no annihilation
process for these ``dressed" states. The states will still be labelled by
the positions of the original unperturbed particles; but now they will
contain admixtures of other particle states at nearby locations.

At any finite order in perturbation theory, we may assume that the
effective Hamiltonian will remain ``local" (that is, interactions
between states will not extend beyond a finite range); and will have the
same bulk symmetries as the original Hamiltonian, such as translation
symmetry. These properties are sufficient to admit a linked cluster approach 
to the calculation of eigenvalues.

We note that the solution of the equations above is not nearly as
efficient as the similarity transformation of Gelfand: in particular,
the solution of the equation (\ref{eq1.14}) is expensive in CPU time and 
memory. 
In the Appendix, we discuss an alternative `2-block' scheme which has the
same efficiency as Gelfand's; but which does not always allow a
successful cluster expansion. 

\subsection{Linked Cluster Expansions}

Let us briefly summarize the linked cluster
approach in various sectors.

\subsubsection{Ground-state energy \label{subsec1}}
\label{subsection2B1}

	The ground-state energy $E_{0}$ is a simple extensive quantity,
and obeys the ``cluster addition property": if C is a cluster (or set of
sites and bonds on the lattice) which is composed of two disconnected
sub-clusters A and B, then
\begin{equation}
                E_{0}^{C} = E_{0}^{A} + E_{0}^{B} \;.
\end{equation}
Hence one finds\cite{nickel,marland,irving1,he,singh1,gelfand1}
that the ground-state energy per site for the bulk lattice can be
expressed purely in terms of contributions from connected sub-clusters
$\alpha$:
\begin{equation}
\epsilon_{0} = \sum_{\alpha} l_{\alpha}
\epsilon_{\alpha} \;,
\label{3}
\end{equation}
where $l_{\alpha}$ is the ``lattice constant", or number of ways per site
that cluster $\alpha$ can be embedded in the bulk lattice, and
$\epsilon_{\alpha}$ is the ``proper energy" or ``cumulant energy" for
the cluster $\alpha$. In the language of Feynman diagrams,
$\epsilon_{\alpha}$ can be thought of as the sum of all connected
diagrams spanning the cluster $\alpha$\cite{hamer2}.

	A similar formula holds for the ground-state energy of any
connected cluster $\alpha$ with open boundaries:
\begin{equation}
E_{0}^{\alpha} = \sum_{\beta} C_{\beta}^{\alpha}
\epsilon_{\beta} \;,
\label{4}
\end{equation}
where $C_{\beta}^{\alpha}$ is the embedding constant of the connected
sub-cluster $\beta$ within cluster $\alpha$.

	Equations (\ref{3}) and (\ref{4}) form the basis for a simple
and efficient recursive algorithm to generate a perturbation series for
$\epsilon_{0}$. The steps are:
\begin{itemize}
\item[i)] Generate a list of clusters $\alpha$, with their lattice
constants $l_{\alpha}$ and embedding constants $C_{\beta}^{\alpha}$,
appropriate to the problem at hand \cite{martin,jaanwork,bonnsite};
\item[ii)] For each cluster $\alpha$, 
the diagonal entry in the 0-particle sector of $H^{\rm eff}$ gives a
perturbation series for the energy 
 $E_{0}^{\alpha}$;
\item[iii)] Now invert equations (\ref{4}) to solve for the cumulant
energies $\epsilon_{\alpha}$, and substitute in (\ref{3}) to obtain the
desired perturbation series for $\epsilon_{0}$.
\end{itemize} 

\subsubsection{1-particle excited states}

\begin{figure}
 \begin{center} 
 \vskip 0cm
 \epsfig{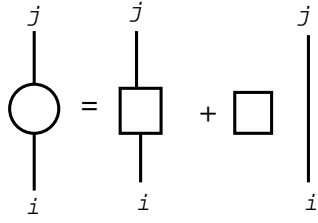}
  \vskip 0cm
 \caption[]
         {Decomposition of a 1-particle matrix element
into irreducible components. The round box denotes the full matrix
element, the square boxes the irreducible matrix elements, and the single
line denotes a delta function.}
 \label{fig2} 
 \end{center}   
\end{figure}

	Gelfand\cite{gelfand2} discovered how to generalize the approach
above to one-particle excited states.
Let
\begin{equation}
E_{1}({\bf i,j}) = \langle {\bf j}| H^{\rm eff}|{\bf i}\rangle
\end{equation}
be the matrix element of $H^{\rm eff}$ between initial 1-particle state $|
{\bf i} \rangle$ and final 1-particle state $|{\bf j}\rangle$,
labelled according to their positions on the lattice. The excited state
energy is not extensive, and does not obey the cluster addition
property; but there is a related quantity which does. If cluster C is
made up of disconnected sub-clusters A and B, and states
$|{\bf i} \rangle$ and $|{\bf j}\rangle$
reside (say) on cluster A, then
\begin{equation}
E_{1}^{C}({\bf i,j}) = E_{1}^{A}({\bf i,j}) + E_{0}^{B} \;.
\end{equation}
But if we define the ``irreducible" 1-particle matrix element
(Fig. \ref{fig2})
\begin{equation}
\Delta_{1}({\bf i,j}) = E_{1}({\bf i,j}) - E_{0} \delta_{{\bf i,j}} \;,
\end{equation}
then
\begin{equation}
\Delta_{1}^{C}({\bf i,j}) = \Delta_{1}^{A}({\bf i,j}) \;, 
\end{equation}
whereas if
$|{\bf i} \rangle$ and $|{\bf j}\rangle$
reside on cluster B, then
\begin{equation}
\Delta_{1}^{C}({\bf i,j}) = \Delta_{1}^{B}({\bf i,j})
\end{equation}
or in general
\begin{equation}
\Delta_{1}^{C}({\bf i,j}) = \Delta_{1}^{A}({\bf i,j}) + \Delta_{1}^{B}
({\bf i,j}) \;,
\label{10}
\end{equation}
where $\Delta_{1}({\bf i,j})$ {\it vanishes} for any cluster not
containing {\bf i} and {\bf j}. Note that a 1-particle state cannot
annihilate from one sub-cluster and reappear on the other, after the
initial block diagonalization.

	From the cluster addition property (\ref{10})
it follows that the elements $\Delta_{1}({\bf i,j})$ can be expanded in
terms of contributions from {\it connected} clusters alone, which are
also ``rooted", or connected to the positions
{\bf i} and {\bf j}. Hence they can be calculated efficiently by an
algorithm like that of subsection (\ref{subsec1}).

\subsubsection{2-particle states}
\begin{figure}[ht]
 \begin{center}
 \epsfig{file=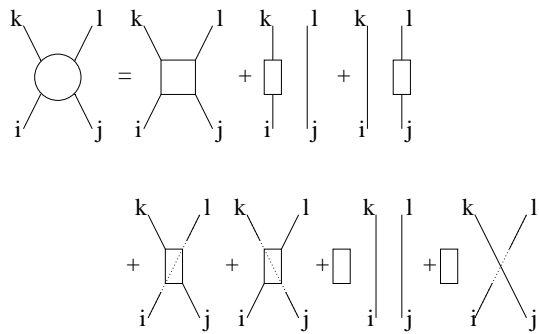, width=7cm}
 \caption[]
         {Decomposition of  two identical particle matrix element
into irreducible components. Notation as in Fig. 2.}
 \label{fig3} 
 \end{center}   
\end{figure}

	The generalization to two-particle states is now not hard to
find. Let
\begin{equation}
E_{2}({\bf i,j;k,l}) = \langle {\bf k,l} | H^{\rm eff} | {\bf i,j}
\rangle
\end{equation}
be the matrix element between initial 2-particle state $ | {\bf i,j}
\rangle$ and final state $ | {\bf k,l} \rangle$. To obtain a
quantity obeying the cluster addition property, we must subtract the
ground-state energy and 1-particle contributions, to form the
irreducible 2-particle matrix element [Fig. \ref{fig3}]:
\begin{eqnarray}
\Delta_{2}({\bf i,j;k,l}) &  = & E_{2}({\bf i,j;k,l}) -E_{0}(\delta_{{\bf
i,k}}\delta_{{\bf j,l}} + \delta_{{\bf i,l}}\delta_{{\bf j,k}}) -
\Delta_{1}({\bf i,k})\delta_{{\bf j,l}} - \Delta_{1}({\bf i,l})
\delta_{{\bf j,k}} \nonumber\\
 & & - \Delta_{1}({\bf j,k}) \delta_{{\bf i,l}} -
\Delta_{1}({\bf j,l}) \delta_{{\bf i,k}} \;.
\end{eqnarray}
This quantity is easily found to be {\it zero} for any cluster
unless {\bf i, j, k} and {\bf l} are all included in that cluster, and
it obeys the cluster addition property. Once again, the block
diagonalization ensures that two particles cannot ``annihilate" from one
cluster and ``reappear" on another disconnected one. Thus the matrix
elements of $\Delta_{2}$ can be expanded in terms of connected clusters
alone, which are rooted or connected to all four positions 
{\bf i, j, k, l}.

\subsection{Calculation of eigenvalues}

	For the ground state energy, a perturbation series for the
eigenvalue was already obtained at the end of subsection
\ref{subsection2B1}. For the excited state sectors, some further work
is required.

\subsubsection{1-particle states}

	A perturbation series for the dispersion relation of the
1-particle states can
be calculated by a Fourier transform. Translation invariance implies
that 
\begin{equation}
\Delta_{1}({\bf i,j}) \equiv \Delta_{1}({\bf \delta}) \;,
\end{equation}
where ${\bf \delta}$ is the difference between positions {\bf i} and
{\bf j}; and that the 1-particle states are eigenstates of momentum:
\begin{equation}
| {\bf K} \rangle = \frac{1}{\sqrt{N}}\sum_{{\bf j}}
\exp(i{\bf K\cdot j}) | {\bf j} \rangle \;,
\end{equation}
(where $N$ is the number of sites in the lattice), with energy gap
\begin{equation}
\omega_{1}({\bf K}) = \sum_{{\bf \delta}} \Delta_{1}({\bf
\delta})\cos({\bf K\cdot\delta}) \;.
\end{equation}
Here we have assumed that $\Delta_{1}({\bf \delta})$ is reflection
symmetric, so that
\begin{equation}
 \Delta_{1}({\bf
- \delta}) = \Delta_{1}({\bf \delta}) \;.
\end{equation}

\subsubsection{2-particle states}
	The calculation of the eigenvalues in this case is a little more
involved than in the 1-particle case. We follow the procedure of
Mattis\cite{mattis}. 

Consider an unsymmetrized state of non-identical
particles, types $a$ and $b$. Then there are $N(N-1)$ states on an $N$-site
lattice, labelled by positions $|{\bf i,j} \rangle$, where {\bf i,j} refer to the
positions of particles $a$ and $b$, respectively.
We have assumed here that two particles may not reside at the same
position (the results are easily amended if this is not the case).
Then the irreducible 2-particle matrix element is:
\begin{eqnarray}
\Delta_{2}^{ab}({\bf i,j;k,l}) &  = & E_{2}^{ab}({\bf i,j;k,l}) 
 \nonumber\\
 & &  
-E_{0}\delta_{{\bf i,k}}\delta_{{\bf j,l}} 
-\Delta_{1}^{a}({\bf i,k})\delta_{{\bf j,l}} 
-\Delta_{1}^{b}({\bf j,l}) \delta_{{\bf i,k}} \;.
\end{eqnarray}
Let the 2-particle eigenstate be: 
\begin{equation}
| \psi \rangle = \sum_{{\bf i,j}} f_{{\bf ij}} | {\bf
i,j} \rangle, \hspace{3mm} (i \neq j) \;, 
\end{equation}
substitute in the Schr\"odinger equation
\begin{equation}
H^{\rm eff} | \psi \rangle = E | \psi \rangle
\end{equation}
and take the overlap with $\langle {\bf i,j} |$, then one obtains:
\begin{equation}
(E -E_{0})f_{{\bf ij}} - \sum_{{\bf k \ne j}} \Delta^{a}_{1}({\bf
k,i})f_{{\bf kj}} - \sum_{{\bf k \ne i}} \Delta^{b}_{1}({\bf k,j})f_{{\bf
ik}} = \sum_{{\bf k,l}} \Delta^{ab}_{2}({\bf k,l;i,j})f_{{\bf kl}} \;, 
\hspace{2mm} ({\bf i \ne j}) \;.
\end{equation}
Completing the sums on the left-hand side, one obtains:
\begin{eqnarray}
(E - E_{0})f_{{\bf ij}} - \sum_{\bf k}[ \Delta^{a}_{1}({\bf
k,i})f_{{\bf kj}} +  \Delta^{b}_{1}({\bf k,j})f_{{\bf
ik}}] & = & \sum_{{\bf k,l}} \Delta^{a,b}_{2}({\bf k,l;i,j})f_{{\bf kl}} \nonumber\\
 & & - \Delta^{a}_{1}({\bf j,i})f_{{\bf jj}} - \Delta^{b}_{1}({\bf i,j})f_{{\bf ii}}
\;,
\label{30}
\end{eqnarray}
The fictitious amplitudes $f_{{\bf ii}}$ are introduced to simplify the
calculations, and are taken to be {\it defined} by these
equations\cite{mattis}. 

	Now define a centre-of-mass position co-ordinate
\begin{equation}
{\bf R} = \frac{1}{4} ({\bf i+j+k+l}) \;,
\end{equation}
and relative co-ordinates
\begin{equation}
{\bf r} = \frac{1}{2} ({\bf i+j-k-l}) \;,
\end{equation}
\begin{equation}
\delta_{1} ={\bf i-j} \;,
\end{equation}
\begin{equation}
\delta_{2} ={\bf k-l} \;.
\end{equation}
Translation invariance then implies that
\begin{equation}
\Delta_{2}({\bf i,j;k,l}) \equiv \Delta_{2}({\bf r,\delta_{1},\delta_{2}}) \;.
\end{equation}

Next, perform a Fourier transformation,
\begin{equation}
f({\bf K,q}) = \frac{1}{N} \sum_{{\bf i,j}} e^{i({\bf k_{1}\cdot i +
k_{2}\cdot j})}f_{{\bf ij}} \;,
\end{equation}
where ${\bf K,q}$ are the centre-of-mass and relative momenta:
\begin{equation}
{\bf K} = ({\bf k}_{1} + {\bf k}_{2}) \;,
\end{equation}
\begin{equation}
{\bf q} = \frac{1}{2}({\bf k}_{1} - {\bf k}_{2}) \;,
\end{equation}
then equation (\ref{30}) leads to
\begin{eqnarray}
[E - E_{0} -\sum_{{\bf \delta}}[ \Delta^{a}_{1}({\bf \delta})\cos({\bf
K\cdot\delta}/2 +{\bf q\cdot\delta})
+ \Delta^{b}_{1}({\bf \delta})\cos({\bf
K\cdot\delta}/2 -{\bf q\cdot\delta})
]f({\bf K,q}) & = & \nonumber\\
  \frac{1}{N} \sum_{{\bf
q'}}f({\bf K,q'})[\sum_{{\bf r,\delta_{1},\delta_{2}}}\Delta^{ab}_{2}
({\bf r,\delta_{1},\delta_{2}})\cos({\bf K\cdot r} + {\bf q\cdot\delta_{1}}
 - {\bf q'\cdot\delta_{2}}) \nonumber\\
 - \sum_{{\bf \delta}}\Delta^{a}_{1}(\delta)\cos({\bf K\cdot\delta}/2
 + {\bf
q \cdot\delta})
  + \Delta^{b}_{1}(\delta)\cos({\bf K\cdot\delta}/2
 - {\bf
q \cdot\delta})
] \;,
\end{eqnarray}
where we have again assumed reflection symmetry
\begin{equation}
\Delta^{a,b}_{1}(\delta) = \Delta^{a,b}_{1}(-\delta) \;,
\end{equation}
\begin{equation}
\Delta^{ab}_{2}({\bf r,\delta_{1},\delta_{2}}) = \Delta^{ab}_{2}({\bf
-r,-\delta_{1},-\delta_{2}}) \;.
\end{equation}

	Finally, look for solutions with definite {\it exchange
symmetry}.

\begin{flushleft}
{\bf Symmetric states}
\end{flushleft}

\begin{equation}
f_{ij} = + f_{ji}
\end{equation}
therefore
\begin{equation}
f({\bf K,q}) = + f({\bf K,-q}) \;.
\end{equation}
`Averaging' over $f({\bf K,\pm q})$ (i.e. taking $\frac{1}{2}[f({\bf K,q})
+ f({\bf K,-q})]$), we get:

\begin{eqnarray}
\{E - E_{0} -\sum_{{\bf \delta}}[ \Delta^{a}_{1}({\bf \delta})
+ \Delta^{b}_{1}({\bf \delta})]\cos({\bf
K\cdot\delta/2})\cos({\bf q\cdot\delta})\}
f({\bf K,q}) & = & \nonumber\\
  \frac{1}{N} \sum_{{\bf
q'}}f({\bf K,q'})[\sum_{{\bf r,\delta_{1},\delta_{2}}}\Delta^{ab}_{2}
({\bf r,\delta_{1},\delta_{2}})\cos({\bf K\cdot r})\cos({\bf q\cdot\delta_{1}})
\cos({\bf q'\cdot\delta_{2}}) \nonumber\\
 - \sum_{{\bf \delta}}[\Delta^{a}_{1}(\delta)
  + \Delta^{b}_{1}(\delta)]\cos({\bf K\cdot\delta}/2)
\cos({\bf
q \cdot\delta})
] \;.
\end{eqnarray}
\begin{flushleft}
{\bf Antisymmetric states}
\end{flushleft}

\begin{equation}
f_{ij} = - f_{ji}
\end{equation}
therefore
\begin{equation}
f({\bf K,q}) = - f({\bf K,-q}) \;.
\end{equation}
`Averaging' over $f({\bf K,\pm q})$ (i.e. taking $\frac{1}{2}[f({\bf K,q})
- f({\bf K,-q})]$), we get:

\begin{eqnarray}
\{E - E_{0} -\sum_{{\bf \delta}}[ \Delta^{a}_{1}({\bf \delta})
+ \Delta^{b}_{1}({\bf \delta})]\cos({\bf
K\cdot\delta/2})\cos({\bf q\cdot\delta})\}
f({\bf K,q}) & = & \nonumber\\
  \frac{1}{N} \sum_{{\bf
q'}}f({\bf K,q'})\sum_{{\bf r,\delta_{1},\delta_{2}}}\Delta^{ab}_{2}
({\bf r,\delta_{1},\delta_{2}})\cos({\bf K\cdot r})\sin({\bf q\cdot\delta_{1}})
\sin({\bf q'\cdot\delta_{2}}) \;.
\end{eqnarray}

\begin{flushleft}
{\bf Identical particles}
\end{flushleft}

If the particles $a$ and $b$ are identical, the solution is the same as for
symmetric states except the labels $a$ and $b$ must now be dropped, and to
avoid double counting it turns out that the $\Delta_{2}$ term must be
multiplied by an extra factor of 1/2:
\begin{eqnarray}
[E - E_{0} -2\sum_{{\bf \delta}} \Delta_{1}({\bf \delta})\cos({\bf
K\cdot\delta}/2)\cos({\bf q\cdot\delta})]f({\bf K,q}) & = & \nonumber\\
  \frac{1}{N} \sum_{{\bf
q'}}f({\bf K,q'})[\frac{1}{2}\sum_{{\bf r,\delta_{1},\delta_{2}}}\Delta_{2}
({\bf r,\delta_{1},\delta_{2}})\cos({\bf K\cdot r})\cos({\bf q\cdot\delta_{1}})
\cos({\bf q'\cdot\delta_{2}}) \nonumber\\
 - 2\sum_{{\bf \delta}}\Delta_{1}(\delta)\cos({\bf K\cdot\delta}/2)
\cos({\bf
q \cdot\delta})] \;.
\end{eqnarray}

The above integral equations can be solved, 
for a given value of ${\bf K}$, using standard discretization
techniques. Instead of 
using continous momentum ${\bf q}$, 
one can use $N$ discretized and equally spaced values of momentum,
so that instead of solving the complicated integral equation, one only 
needs to compute the eigenvalue and
eigenvector of an $N\times N$ matrix for the discretized
system. Notice that the matrix is 
nonsymmetric  due to the unphysical $f_{ii}$ term we have
introduced in Eq. (\ref{30}), but even so the eigenvalues obtained from 
this matrix are  real. The solutions we obtain also
include an unphysical one with eigenvalue equal to 0 (this is also due to
the unphysical $f_{ii}$ term). The results obtained from the calculation with
discretized momenta will converge to those with continous momentum
as $N\to \infty$. Actually for those bound states with finite coherence length,  
the calculation will normally be
well converged for quite small values of $N$, 
but for unbound states, we have an infinite coherence length, so
 one may need to
do finite $N$ extrapolations to get results at $N=\infty$.

There are two methods to compute the eigenvalues of the matrix for the discretized
system.
Obviously one can get numerical results for the eigenvalues, 
for a given value of coupling $\lambda$ and momentum ${\bf K}$,  
 via standard numerical
techniques where we just perform a naive sum for the series in $\Delta_1$ and $\Delta_2$.
The results presented in a preceding Letter \cite{shortpaper} are based on 
this method;
but then one cannot carry out a series extrapolation, and so one
may not be able to reach a region of critical coupling.
A better technique is to compute the series in $\lambda$ for the eigenvalues 
through degenerate perturbation theory: that is,
by explicit diagonalization of the matrix within the degenerate subspace, order
by order in perturbation theory,
 and then one can perform a
 series extrapolation. The problem with this method is that the series does not
always exist, for example for those bound states appearing at some nonzero value of $\lambda$.

The two particle continuum is delimited by the maximum (minimum)  energy of
two single particle excitations whose combined momentum is the center
of mass momentum. Apart from the unphysical eigenvalue, 
there may be multiple solutions above/below the two-particle continuum. 
Those solutions with energy
below the bottom edge of the continuum are the bound states, while the solutions with energy
higher than the upper edge of the continuum are the antibound states. The binding energy
is defined as the energy difference
between the lower edge of the continuum and the energy of the bound state,
while the antibinding energy is defined as the energy difference
between the upper edge of  continuum and the energy of an antibound state.

Note that the series for $\Delta_2$ may depend on the transformation used to block diagonalize
the Hamiltonian. If we compute $\Delta_2$ (and also $\Delta_1$) to order $n$,
the resulting series for the 2-particle energy obtained from the above integral equation
will have two parts: the part up to order $n$ is independent of the transformation,
while the higher order terms are incomplete, and may depend on the transformation. 
The numerical solution of the integral equation may also
depend partly on the transformation, since it contains the higher order term. Also
note that the series for $\Delta_2$ need not have
any singularities. The singularities, if they exist, arise in the solution of
the Schr\"odinger equation, so our method should be able to explore
new bound states arising as we vary the momentum ${\bf K}$. If we get a numerical solution,
rather than a series solution, to the Schr\"odinger equation,  
we should also be able to
explore new bound states arising as we increase $\lambda$ 
as long as the naive sum to the series converges.

\subsection{Finite Lattice Approach}
\label{finite-lattice}

Once the cluster expansions for the irreducible matrix elements $\Delta_1$
and $\Delta_2$ have been developed, the Schr\"odinger equation in the
two-particle subspace can be solved by an alternative method that
works in coordinate space rather than momentum space. By restricting
to a finite but large system with periodic boundary conditions, the
two-particle Schr\"odinger equation becomes a finite-dimensional matrix
of equations. The cluster expansion results provide the matrix elements
of the effective Hamiltonian as a power series in the expansion parameter.
The centre of mass momentum is a conserved quantity,
thus, for a given value of the centre of mass
momentum, one is left with a Schr\"odinger equation in the separation
variable. One can truncate the perturbation theory at a given order
and solve the Schr\"odinger equation numerically. One can then vary
the size of the system, which only increases the dimension of matrix
to be diagonalized linearly, to study convergence. We have frequently used this
method to compare with and check the momentum-space discretization solutions.

This `finite lattice approach' also allows us to obtain power series
expansions for bound state energies, by a non-degenerate perturbation
theory, provided the bound state exists ``localized" in the limit $\lambda\to 0$. 
For those ``extended" bound states in the limit $\lambda\to 0$\cite{longj12d}, one still needs
to do degenerate perturbation theory, just as in the case of the momentum-space 
discretization solutions. Another advantage of this method over the the momentum-space 
discretization technique is that the matrix one deals with is always symmetric.

Furthermore, it gives us explicit real-space wave functions, from
which the coherence length and other properties can be deduced.
The coherence length $L$ is defined by
\be
L = {\sum_{{\bf r}} \vert {\bf r} \vert f_{{\bf r}}^2\over \sum_{{\bf r}} f_{{\bf r}}^2}
\ee
where $f_{{\bf r}}$ is the amplitude (the eigenvector) for two single-particle excitations
 separated by distance ${\bf r}$.

\section{RESULTS}

We apply the new method to the (1+1)D transverse Ising model and a two-leg
spin-$\case 1/2$ Heisenberg ladder.

\subsection{Transverse Ising model}
\label{secIsing}
In order to verify that our new technique is giving the correct results, 
we firstly apply it to a simple model, the $S=\case 1/2$ transverse Ising model
in (1+1)-dimensions, which is exactly solvable in terms of free fermions.
The Hamiltonian for it reads
\be
H = \sum_i (1-\sigma^z_i) - \lambda \sum_{i} \sigma^x_i  \sigma^x_{i+1} \;,
\ee
Here we take the first term as the unperturbed Hamiltonian $H_0$,
and the second term as the perturbation $H_1$.
The ground state of $H_0$ is the unique state with all spins pointing up.
The lowest excited states (1-particle excitations) for $H_0$ flip 
one of the spin from spin up
to spin down. The exact result\cite{exactIsing} for the 1-particle
dispersion relation is
\be
E_1 (q) = 2 \sqrt{ 1 + \lambda^2 - 2 \lambda \cos q } \;.
\ee

For the 2-particle excitations, the unperturbed states have two spins down.
Since this model can be mapped into free fermions, 
there are no 2-particle bound states, and the 2-particle excitation energy is simply
the sum of two 1-particle dispersions, that is
\be
E_2(q_1, q_2) = 2 \sqrt{ 1 + \lambda^2 - 2 \lambda \cos q_1 } + 
  2 \sqrt{ 1 + \lambda^2 - 2 \lambda \cos q_2 } \label{eqIsingE2} \;,
\ee
where $q_1$ and $q_2$ are the momenta of each particle. Note that this is a non-trivial
example for our method as the similarity transformation does not even lead to
a cluster expansion.
 
\begin{figure}
 \begin{center} 
 \vskip -0.5cm
 \epsfig{file=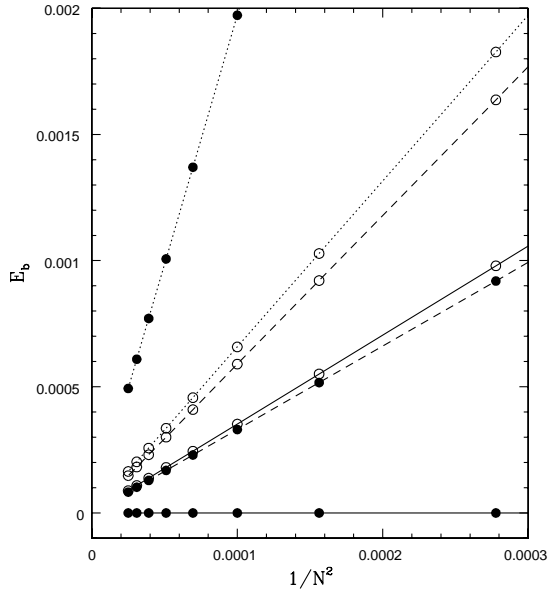, height=10cm}
  \vskip -1.2cm
 \caption[]
         {The binding energy (full points) and antibinding energy (open points)
         $E_b$ versus $1/N^2$ ($N$ is the size of the matrix)
         for the transverse Ising model with coupling $\lambda=0.5$
         and momentum $k=0$ (dotted lines), $\pi/2$ (dashed lines), $\pi$ (solid lines).
          }
 \label{figIsing} 
 \end{center}   
\end{figure}

We have implemented the algorithm described above for this model. For the
1-particle excitation,
we can easily reproduce the exact results 
through the different block diagonalization schemes 
mentioned before. For the 2-particle excitations, although there are no bound states,
the terms $\Delta_2$ are not zero. That is because we are using the spin 
representation; in a fermion representation, the $\Delta_2$ would be expected
to vanish.
We have computed them to order $\lambda^{12}$ by using the 2-block method.
The series coefficients up to order $\lambda^6$ are given\cite{wwwsite} in 
Table \ref{tabIsing}.
With these series, one can solve the discretized version of the integral 
equation to get the binding and antibinding energy for any given value of 
momentum $k$ and coupling $\lambda$.
Our results show that for all $k$ and $\lambda$, the binding/antibinding energy
scales as $1/N^2$, and approaches to zero as $N\to \infty$: this is consistent 
with the absence of bound/antibound states in this model. 
The results for $\lambda=0.5$ and $k=0,\pi/2,\pi$ are shown in Fig. 
\ref{figIsing}.  
We have also checked that the resulting series for $E_2$ agrees with 
(\ref{eqIsingE2}) for the lowest and highest energy of 2-particle states up to 
12th order, and the coherence length is infinity, as expected.

\subsection{Heisenberg ladder}
\label{secLadder}


The second model we have investigated is 
the 2-leg spin-$\frac{1}{2}$ Heisenberg ladder, where the Hamiltonian is
\begin{equation}
H = \sum_i \{ J_{\perp} {\bf S}_i\cdot {\bf S}^{\prime}_{i}
+ J [ {\bf S}_i\cdot {\bf S}_{i+1} 
        + {\bf S}_i^{\prime}\cdot {\bf S}^{\prime}_{i+1}
         ] \} \;,
\label{H:ladder}
\end{equation}
where ${\bf S}_i$ (${\bf S}^{\prime}_{i}$) denotes the  spin at site $i$ of 
the first (second) chain. $J$
is the interaction between nearest-neighbor spins along the
chain, and $J_{\perp}$ is the interaction between nearest-neighbor spins
along the rungs. 
In the present paper the intra-chain coupling is taken to be antiferromagnetic (that is,
$J_{\perp} > 0$ ) whereas the interchain coupling $J$ can be either antiferromagnetic or ferromagnetic. 

The antiferromagnetic Heisenberg ladder has attracted a good deal of attention
recently \cite{Dag96,Gopalan:94,Oitmaa:96,Eder:97,Damle:98,Jurecka:99,Kotov:98,Kotov:99}.
It is of experimental interest in that there are a number of 
quasi-one-dimensional compounds which may be described by the model
\cite{Dag96}. 
It is also a prime example of a one-dimensional antiferromagnetic system with a
gapped excitation spectrum.
Damle and Sachdev \cite{Damle:98} as well as Sushkov and Kotov \cite{Kotov:98} 
have shown that the system exhibits two-particle bound states, one singlet and
one triplet. 
Our aim here is to explore the properties of these bound states more closely. 

In the dimer limit $J/J_{\perp}=0$, 
the ground state is the product state with the spins
on each rung forming a spin singlet. 
The first excited state consists of a spin triplet excitation on one of the
rungs. As $J/J_{\perp}$ increases, this state
evolves smoothly, and the system has a
gapped excitation spectrum\cite{Gopalan:94,Oitmaa:96,Eder:97}.  
The dimer expansions have been computed  previously up to order 
$(J/J_{\perp})^{23}$ for the ground-state energy
and up to order $(J/J_{\perp})^{13}$ for the 1-particle triplet excitation
spectrum\cite{Oitmaa:96}.
The occurrence of two-particle
bound states in this model has been shown by first-order
strong-coupling expansions \cite{Damle:98,Jurecka:99} as well as a
leading order calculation using the analytic Brueckner
approach\cite{Kotov:98,Kotov:99}.

\begin{figure}
 \begin{center} 
 \vskip -0.5cm
 \epsfig{file=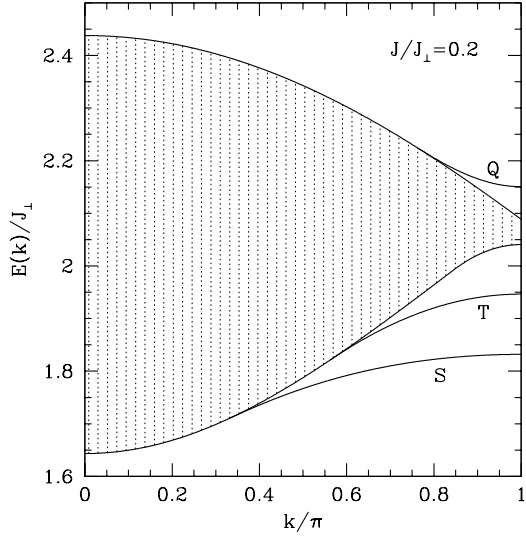, height=10cm}
  \vskip -1.2cm
 \caption[]
         {The excitation spectrum for the Heisenberg spin ladder at
          $J/J_{\perp} = 0.2$. Beside the two-particle continuum (gray 
          shaded), there are
          three massive quasiparticles: a singlet bound state (S), 
          a triplet bound state (T) below the continuum and a
          quintet antibound state (Q) above the continuum.
          }
 \label{figLaddermk} 
 \end{center}   
\end{figure}

\begin{figure}
 \begin{center} 
 \vskip -0.5cm
 \epsfig{file=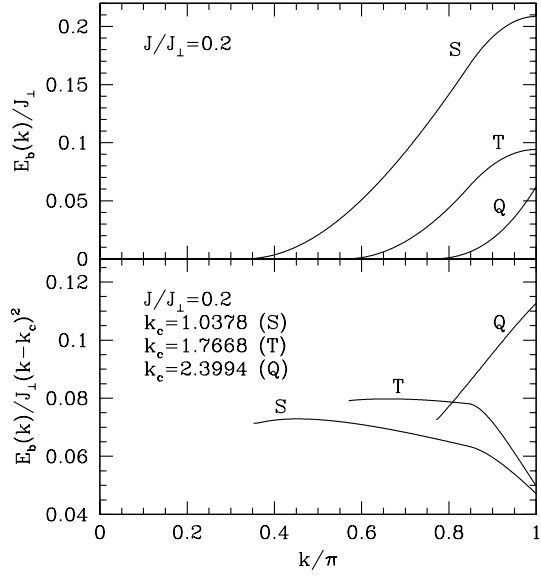, height=10cm}
  \vskip -1.2cm
 \caption[]
         {The binding/antibinding energy $E_b/J_{\perp}$ 
         (upper window) and the rescaled binding/antibinding energy 
         $E_b/J_{\perp}(k-k_c)^2$ (lower window) for the Heisenberg spin ladder at
          $J/J_{\perp} = 0.2$. 
          }
 \label{figLadderEb} 
 \end{center}   
\end{figure}

Here we have calculated series for the dispersions of the 2-particle bound states
up to order $(J/J_{\perp})^7$ for the singlet bound state ($S$), and
to order $(J/J_{\perp})^{12}$ for the triplet bound state ($T$) and the quintet 
antibound state ($Q$). 
The reason why the singlet series is computed to only 7th order compared
to 12th order for the triplet and quintet states is that the
singlet has the same quantum numbers as the ground state. Thus
a much more elaborate orthogonalization method is required to implement the
cluster expansion for the singlet. For the triplet and quintet bound states,
we can use the similarity transformation or the 2-block 
orthogonal transformation to implement the cluster expansion.
Up to order $(J/J_{\perp})^3$, the dispersion for the
singlet bound state ($E_S/J_{\perp}$), triplet bound state ($E_T/J_{\perp}$), 
quintet antibound state ($E_Q/J_{\perp}$)  are
\begin{mathletters}
\label{eqLadder2p}
\bea
E_S/J_{\perp} &=& 2 - \frac{3\,x}{2} + \frac{19\,x^2}{16} + \frac{9\,x^3}{32} + 
  \left( - \frac{x}{2} + \frac{x^2}{8} - \frac{51\,x^3}{128} \right) \,\cos (k) \nonumber \\
  && +   \left( - \frac{5\,x^2}{16} - \frac{21\,x^3}{32} \right) \,\cos (2\,k) - \frac{37\,x^3\,\cos (3\,k)}{128}
   + O(x^4) \label{eqEs} \;, \\
E_T/J_{\perp} &=&  2 - \frac{3\,x}{2} + \frac{11\,x^2}{8} + \frac{17\,x^3}{16} 
+ \left( -x - \frac{x^2}{4} + \frac{9\,x^3}{16} \right) \,\cos (k) \nonumber \\
&& +   \left( - \frac{x^2}{2} - \frac{x^3}{2} \right) \,\cos (2\,k)
 - \frac{5\,x^3\,\cos (3\,k)}{16} + O(x^4)  \label{eqEt} \;, \\
E_Q/J_{\perp} &=&  2 + \frac{3\,x}{2} + \frac{11\,x^2}{8} - \frac{3\,x^3}{16} 
+ \left( x - \frac{x^2}{4} - \frac{27\,x^3}{16} \right) \,\cos (k) \nonumber \\
&& + \left( - \frac{x^2}{2} - \frac{3\,x^3}{8} \right) \,\cos (2\,k) + \frac{7\,x^3\,\cos (3\,k)}{16}  + O(x^4) \label{eqEq} \;,
\eea
\end{mathletters}
where $x\equiv J/J_{\perp}$.
The full dispersion series for the singlet state, and the series for the energy gap at $k=\pi$
for singlet, triplet and quintet bound/antibound states and the lower edge and upper edge of continuum are listed in
Table \ref{tabLadder} and \ref{tabserKpi};   the other series  are
available upon request\cite{wwwsite}.
Figs. \ref{figLaddermk} and  \ref{figLadderEb}
show the dispersion and the binding/antibinding energy 
at $J/J_{\perp}=0.2$ for the two-particle continuum as well as the
the two-particle bound/antibound states. Here we can see
there is a singlet ($S=0$) and a triplet ($S=1$) 
bound state of two elementary triplets below
the two-particle continuum, and a quintet ($S=2$) antibound state
above the continuum. 

From these graphs, we can also see
that these bound/antibound states exist only when the momentum $k$ 
is larger than some ``critical momentum" $k_c$:  the series in Eq. (\ref{eqLadder2p})
and Table \ref{tabLadder} are valid only for $k\geq k_c$.
It is interesting to explore the behaviour of the binding energies
near this critical momentum.
From the series for the one-particle and two-particle dispersions,
one can get leading order results for $k_c$ as
\be
k_c = \left\{ \begin{array}{ll}  
\sqrt{10 x}  + O(x)                                         & \mbox{$S=0$} \\
2 \pi/3 - 5 x/(2 \sqrt{3}) - 109 x^2/(48 \sqrt{3}) + O(x^3) & \mbox{$S=1$} \\
2 \pi/3 + 5 x/(2 \sqrt{3}) + 47 x^2/(48 \sqrt{3}) + O(x^3)  & \mbox{$S=2$}
\end{array}
\right.
\label{eqladderKc}
\ee
and in the limit $k\to k_c$, the behaviour of the binding energy is
\bea
E_b/J &=& (k-k_c)^2 [5 x/8 + 975 x^2/128+ O(x^3) ] \nonumber \\
&& + (k-k_c)^3 [ 12+115 x + O(x^2)] \sqrt{10 x}/192 + O[(k-k_c)^4]
\eea
for the singlet bound state, and
\bea
E_b/J &=&  
(k-k_c)^2 [3/8 - x/32 + 0.45313 x^2 + O(x^3)] \nonumber \\
&&+ (k-k_c)^3 [\sqrt{3}/16 - 53x/(64\sqrt{3}) + 0.19245 x^2 + O(x^3) ]  + O[(k-k_c)^4]
\eea  
for the triplet bound state. For the quintet antibound state, the antibinding energy is
\bea
E_b/J &=&
(k-k_c)^2 [3/8 + x/32 - 0.40625 x^2 + O(x^3) ] \nonumber \\
&&+(k-k_c)^3 [\sqrt{3}/16 + 53 x /(64\sqrt{3}) + 1.00886 x^2 + O(x^3)] + O[(k-k_c)^4] \;.
\eea
Here one can see that for all bound/antibound states the ``critical index" is 2, 
independent of the order of expansion, so one expects that this is {\it exact}.

\begin{figure}
 \begin{center} 
 \vskip -0.5cm
 \epsfig{file=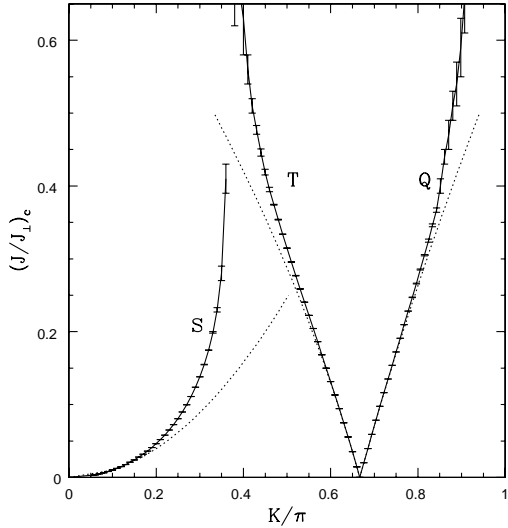, height=10cm}
  \vskip -1.2cm
 \caption[]
         {The critical $(J_/J_{\perp})_c$ versus $k$ for singlet (S), triplet (T) and quintet (Q)
         bound/antibound states of the Heisenberg ladder. The solid lines with errorbars
         are the results obtained from Dlog Pad\'e approximants to the series for binding/antibinding
         energy $E_b$ for given $k$, while the dotted lines are the
         results of Eq. (\ref{eqladderKc}). }
 \label{figLadderPD} 
 \end{center}   
\end{figure}

A better way to locate the critical line in the $(J/J_{\perp})$-$k$ plane is to calculate
the Dlog Pad\'e approximants to the series for the
binding/antibinding energy at a fixed momentum $k$.
For those critical points lying at $x_c < 0.2$,
the resulting  critical point and critical index 
are very accurate, correct up to 5 digits, 
and again one finds the critical index is exactly 2.
The results for the triplet bound state at $k=3\pi/5$ are given in Table
\ref{tabLadderPade}.
The results for the critical points are given in Fig. \ref{figLadderPD}, together
with the results from Eq. (\ref{eqladderKc}).
From this figure, one can see that as $J/J_{\perp}\to \infty$, $k_c$ for the singlet and
triplet bound states approaches the same value, about $0.4\pi$, while $k_c$ for
the quintet antibound state approaches $\pi$. 
To demonstrate that $E_b$ is proportional to $(k-k_c)^2$ near $k_c$,
we also plot in Fig. \ref{figLaddermk} the results for 
$E_b/J_{\perp}(k-k_c)^2$ at $J/J_{\perp}=0.2$.

\begin{figure}
\begin{center} 
\vskip -0.5cm
\epsfig{file=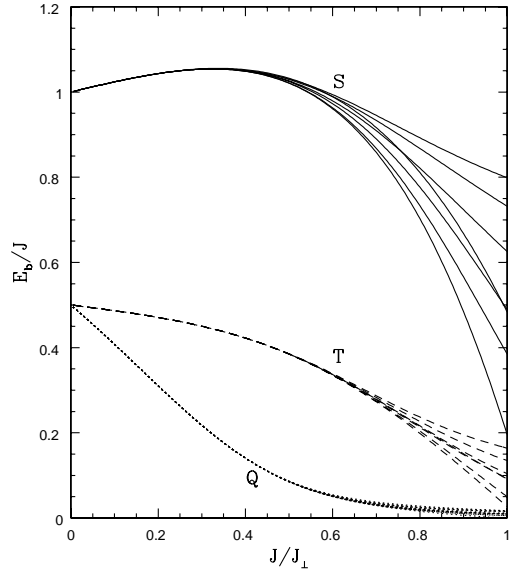, height=10cm}
\vskip -1.2cm
\caption[]
{The binding/antibinding energy $E_b$ at $k=\pi$ versus  
          $J/J_{\perp}$ for singlet ($S$), triplet ($T$) and quintet (Q)
          bound/antibound
         states of the Heisenberg ladder.
         Several different integrated differential approximants to
the series are shown.}
\label{figLadderEb_kpi} 
\end{center}
\end{figure}

The binding/antibinding energy at  $k=\pi$ for bound/antibound states  versus $J/J_{\perp}$ 
is plotted in 
Fig. \ref{figLadderEb_kpi}. In the limit $J/J_{\perp}\to 0$, $E_b/J_{\perp}$ is
proportional to $J/J_{\perp}$, so in the figure we plot $E_b/J$ versus $J/J_{\perp}$. 
We can see that as $J/J_{\perp}$ increases, $E_b/J$ for the singlet bound state firstly increases,
passes through a maximum at about
 $J/J_{\perp}=0.4$, then decreases, while
$E_b/J$ for the triplet bound state and the quintet antibound state decreases monotonically.
At $J/J_{\perp} = 1/2$, 
we find the binding/antibinding energies at $k=\pi$
for the singlet, triplet, and quintet bound/antibound
states are $E_b/J=$1.03(3), 0.385(1) and 0.0855(5), respectively.
The binding energy for the singlet bound state is substantially larger than 
the value 0.70 obtained in \cite{Kotov:99}.

\begin{figure}
 \begin{center} 
 \vskip -0.5cm
 \epsfig{file=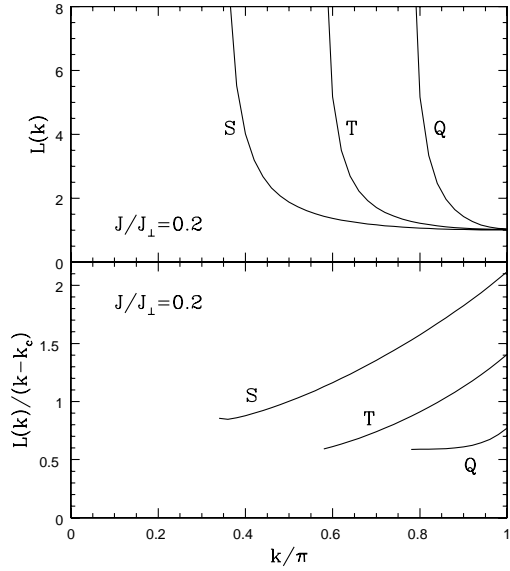, height=10cm}
 \vskip -1.2cm
 \caption[]
          {The coherence length $L$ versus momentum $k$ for singlet (S), 
          triplet (T) and quintet (Q) bound/antibound states of the Heisenberg ladder
          at $J/J_{\perp}=0.2$.}
 \label{figLadderCL_xp2} 
 \end{center}
\end{figure}

We have also computed the coherence length $L$ for these bound/antibound states. 
The results for $J/J_{\perp}=0.2$ are shown
in Fig. {\ref{figLadderCL_xp2}, where we find that $L$ diverges as $1/(k-k_c)$ as $k$ approaches
$k_c$. 
This is to be expected, as the state becomes unbound at that point.
The  coherence length at $k=\pi$ versus $J/J_{\perp}$ is shown in Fig. \ref{figLadderCL_kpi},
where we can see that at $J=0$,  $L=1$. This is as expected,
as the formation of these bound 
states is due to the attraction  of two triplets on neighboring sites. 
As $J/J_{\perp}$ increases, the coherence length $L$ increases slowly.
$L$ for the quintet antibound state is larger than that for the triplet bound state, 
which is larger than for the singlet bound state.

\begin{figure}
 \begin{center} 
 \vskip -0.5cm
 \epsfig{file=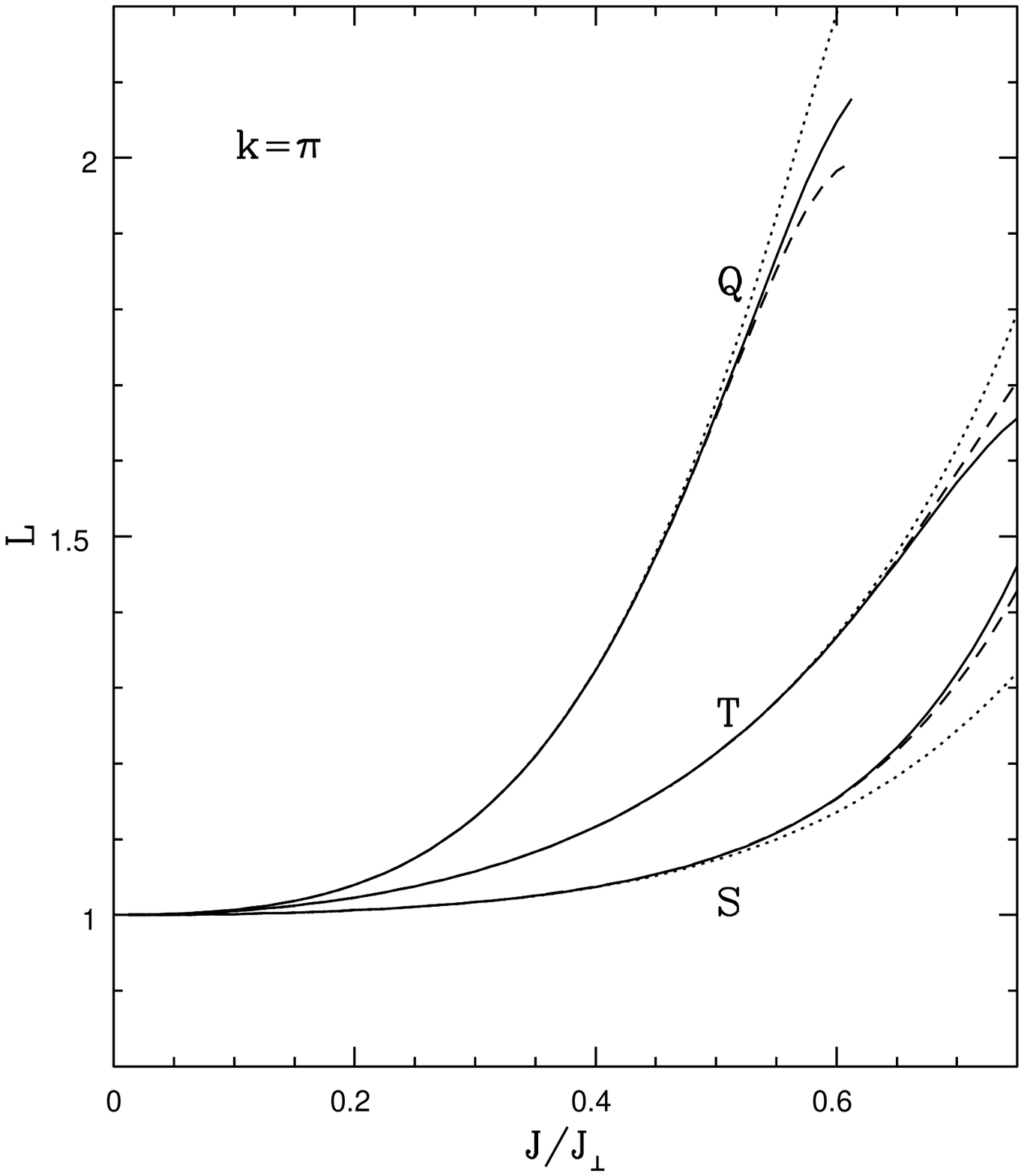, height=10cm}
 \vskip -1.2cm
 \caption[]
          {The coherence length $L$ versus $J/J_{\perp}$ at $k=\pi$ for singlet (S), 
          triplet (T) and quintet (Q) bound/antibound states of the Heisenberg ladder.}
 \label{figLadderCL_kpi} 
 \end{center}
\end{figure}

\section{Conclusions}

In conclusion, we have developed new strong-coupling expansion methods to study
two-particle spectra of quantum lattice models. 
We described in full detail the block diagonalization of the Hamiltonian,
order by order in perturbation theory, to
construct an effective Hamiltonian in the two-particle subspace. 
This work is closely related to Gelfand's prior work on using similarity
transformations to obtain an effective Hamiltonian for the single-particle 
subspace \cite{gelfand2}. We found that one needs to define a
two-particle irreducible matrix element, for a cluster expansion to
exist. Furthermore, one needs to maintain explicit
orthogonality in the transformations in order to study the two-particle
subspace characterized by identical quantum numbers to the ground state.
An example of the latter is the two-particle
singlet excitation sector in dimerized spin models.

We have discussed the solution of the integral equation one obtains by a 
Fourier transformation of the two-particle Schr\"odinger equation 
and by a `finite-lattice approach'.
These allow us to precisely determine the low-lying excitation spectra of the 
models at hand, including all two-particle bound/antibound states.
Furthermore, we have shown that one can generate series expansions
directly for the dispersions
of the bound/antibound states, provided these bound states exist in
the limit $\lambda\to 0$. These allow us to apply series 
extrapolation techniques such as Dlog Pad\'es and differential approximants
to study binding energies even when the perturbation parameter is not small.

We applied the method to the (1+1)D transverse Ising model and the two-leg 
spin-$\case 1/2$ Heisenberg ladder. While the first model does not include any
bound states, we find a singlet and a triplet bound state in the latter model
as well as a quintet antibound state. We generated explicit expressions for
the dispersions of these states as series in the exchange couplings.
Further, we have determined the critical momenta $k_c$, where these
additional massive quasi-particles merge with the two-particle continuum, 
which are non-zero for all three states. 
The explicit expressions of the binding energies at the respective critical 
momenta are found to contribute first in order $(k-k_c)^2$, independent of the
order of the strong coupling expansion.
We computed the coherence length for these states and find that the coherence
length diverges as one approaches the critical momentum where these states 
become unbound.

There are several possible direction for future research along these lines.
Of course there are many different models to which these methods might be applied. In particular, it
remains to show that the linked cluster expansion works sucessfully for two or higher
dimensional models. One would also like to know how to calculate other quantities
associated with multipartcle excitations, such as spectral weights,
lifetimes, and scattering S-matrices. The latter would provide a handle on some
important dynamical properties of the system.

\acknowledgments

This work was initiated at the Quantum Magnetism program at the ITP at 
UC Santa Barbara which is supported by US National Science Foundation 
grant PHY94-07194.
The work of ZW and CJH was supported by a grant from the Australian
Research Council: they thank the New South Wales Centre for Parallel Computing
for facilities and assistance with the calculations.
RRPS is supported in part by NSF grant number DMR-9986948.
ST gratefully acknowledges support by the German National Merit
Foundation and Bell Labs, Lucent Technologies.
HM wishes to thank the Yukawa Institute for Theoretical Physics for
hospitality.

\appendix
\section{2-Block Approach}

There is an alternative way to perform the block diagonalization of
Section \ref{secIIA}, which is almost as efficient as Gelfand's
similarity transformation. The idea is to separate the effective
Hamiltonian into only {\it two} blocks, one containing the states in the
sector of interest (e.g. the 1-particle states, or the 2-particle
states), and the other containing all other states.
One can prove in this two-block approach that $O^{n)}$ determined in
this way is antisymmetric with respect to the off-diagonal blocks, and
{\it symmetric} with respect to the diagonal blocks. Rather than use the
complicated equation (\ref{eq1.14}), one can then determine the diagonal
blocks of $O^{n)}$ in a much more efficient way by the orthogonality
condition (\ref{eq1.8}) which can be rewritten in the following form:

\begin{equation}
\{O^{n)} + O^{n)T}\}_{ij} = - \sum_{m=1}^{n-1}\{O^{m)}O^{n-m)T}\}_{ij}
\end{equation}
for elements in the diagonal blocks. Thus one can dispense with the
matrix $S$, and work with $O$ only.

Unfortunately, although it is more efficient, this approach does not
always seem to allow a successful cluster expansion. The reason for this
is not understood at the present time.

\newpage
\setdec 0.00000000000
\begin{table}
\squeezetable
\caption{Series coefficients for 
$\Delta_2(r,\delta_1,\delta_2) = \sum_{k} \Delta_2^k (r,\delta_1,\delta_2)  x^{k} $
in the (1+1)D transverse Ising model, obtained by the 2-block method.
Nonzero coefficients $\Delta_2^k (r,\delta_1,\delta_2) $
up to order $k=6$  are listed.}\label{tabIsing}
\begin{tabular}{rr|rr|rr|rr}
 \multicolumn{1}{c}{$(k,2r,\delta_1,\delta_2)$} &\multicolumn{1}{c|}{$\Delta_2^k (r,\delta_1,\delta_2)/4$}
&\multicolumn{1}{c}{$(k,2r,\delta_1,\delta_2)$} &\multicolumn{1}{c|}{$\Delta_2^k (r,\delta_1,\delta_2)/4$}
&\multicolumn{1}{c}{$(k,2r,\delta_1,\delta_2)$} &\multicolumn{1}{c|}{$\Delta_2^k (r,\delta_1,\delta_2)/4$}
&\multicolumn{1}{c}{$(k,2r,\delta_1,\delta_2)$} &\multicolumn{1}{c}{$\Delta_2^k (r,\delta_1,\delta_2)/4$} \\
\hline
 ( 2,-2, 1, 1) &\dec    5.000000$\times 10^{-1}$ &( 4, 4, 2, 2) &\dec    1.562500$\times 10^{-1}$ &( 5,-5, 3, 2) &\dec    1.093750$\times 10^{-1}$ &( 6, 4, 3, 1) &\dec $-$5.468750$\times 10^{-2}$  \\
 ( 2, 2, 1, 1) &\dec    5.000000$\times 10^{-1}$ &( 4,-4, 3, 1) &\dec    1.562500$\times 10^{-1}$ &( 5, 5, 3, 2) &\dec    1.093750$\times 10^{-1}$ &( 6,-6, 1, 5) &\dec    8.203125$\times 10^{-2}$  \\
 ( 3,-3, 1, 2) &\dec    2.500000$\times 10^{-1}$ &( 4, 4, 3, 1) &\dec    1.562500$\times 10^{-1}$ &( 5,-5, 4, 1) &\dec    1.093750$\times 10^{-1}$ &( 6, 6, 1, 5) &\dec    8.203125$\times 10^{-2}$  \\
 ( 3, 3, 1, 2) &\dec    2.500000$\times 10^{-1}$ &( 5,-3, 1, 2) &\dec $-$7.812500$\times 10^{-2}$ &( 5, 5, 4, 1) &\dec    1.093750$\times 10^{-1}$ &( 6,-6, 2, 4) &\dec    8.203125$\times 10^{-2}$  \\
 ( 3,-3, 2, 1) &\dec    2.500000$\times 10^{-1}$ &( 5, 3, 1, 2) &\dec $-$7.812500$\times 10^{-2}$ &( 6,-2, 1, 1) &\dec $-$1.953125$\times 10^{-2}$ &( 6, 6, 2, 4) &\dec    8.203125$\times 10^{-2}$  \\
 ( 3, 3, 2, 1) &\dec    2.500000$\times 10^{-1}$ &( 5,-3, 2, 1) &\dec $-$7.812500$\times 10^{-2}$ &( 6, 2, 1, 1) &\dec $-$1.953125$\times 10^{-2}$ &( 6,-6, 3, 3) &\dec    8.203125$\times 10^{-2}$  \\
 ( 4,-2, 1, 1) &\dec $-$1.250000$\times 10^{-1}$ &( 5, 3, 2, 1) &\dec $-$7.812500$\times 10^{-2}$ &( 6,-4, 1, 3) &\dec $-$5.468750$\times 10^{-2}$ &( 6, 6, 3, 3) &\dec    8.203125$\times 10^{-2}$  \\
 ( 4, 2, 1, 1) &\dec $-$1.250000$\times 10^{-1}$ &( 5,-5, 1, 4) &\dec    1.093750$\times 10^{-1}$ &( 6, 4, 1, 3) &\dec $-$5.468750$\times 10^{-2}$ &( 6,-6, 4, 2) &\dec    8.203125$\times 10^{-2}$  \\
 ( 4,-4, 1, 3) &\dec    1.562500$\times 10^{-1}$ &( 5, 5, 1, 4) &\dec    1.093750$\times 10^{-1}$ &( 6,-4, 2, 2) &\dec $-$5.468750$\times 10^{-2}$ &( 6, 6, 4, 2) &\dec    8.203125$\times 10^{-2}$  \\
 ( 4, 4, 1, 3) &\dec    1.562500$\times 10^{-1}$ &( 5,-5, 2, 3) &\dec    1.093750$\times 10^{-1}$ &( 6, 4, 2, 2) &\dec $-$5.468750$\times 10^{-2}$ &( 6,-6, 5, 1) &\dec    8.203125$\times 10^{-2}$  \\
 ( 4,-4, 2, 2) &\dec    1.562500$\times 10^{-1}$ &( 5, 5, 2, 3) &\dec    1.093750$\times 10^{-1}$ &( 6,-4, 3, 1) &\dec $-$5.468750$\times 10^{-2}$ &( 6, 6, 5, 1) &\dec    8.203125$\times 10^{-2}$  \\
\end{tabular}
\end{table}

\setdec 0.000000000000
\begin{table}
\squeezetable
\caption{Series coefficients for the dispersion 
$E (k)/J_{\perp} = \sum_{k,n} a_{k,n} x^{k} \cos (n k) $ 
for the singlet bound state of the Heisenberg ladder.
 Nonzero coefficients $a_{k,n}$
up to order $k=7$  
 are listed. Note that the series are valid only for $k\geq k_c$.}\label{tabLadder}
\begin{tabular}{rr|rr|rr|rr}
\multicolumn{1}{c}{(k,n)} &\multicolumn{1}{c|}{$a_{k,n}$}
&\multicolumn{1}{c}{(k,n)} &\multicolumn{1}{c|}{$a_{k,n}$}
& \multicolumn{1}{c}{(k,n)} &\multicolumn{1}{c|}{$a_{k,n}$}
& \multicolumn{1}{c}{(k,n)} &\multicolumn{1}{c}{$a_{k,n}$} \\
\hline
 ( 0, 0) &\dec 2.0000000       &( 2, 1) &\dec 1.2500000$\times 10^{-1}$ &( 5, 2) &\dec $-$3.9255371       &( 5, 4) &\dec $-$1.5305176        \\
 ( 1, 0) &\dec $-$1.5000000       &( 3, 1) &\dec $-$3.9843750$\times 10^{-1}$ &( 6, 2) &\dec $-$1.0420853$\times 10^{1}$ &( 6, 4) &\dec $-$5.0236816        \\
 ( 2, 0) &\dec 1.1875000       &( 4, 1) &\dec $-$1.9453125       &( 7, 2) &\dec $-$2.8697990$\times 10^{1}$ &( 7, 4) &\dec $-$1.5335112$\times 10^{1}$  \\
 ( 3, 0) &\dec 2.8125000$\times 10^{-1}$ &( 5, 1) &\dec $-$5.2039795       &( 3, 3) &\dec $-$2.8906250$\times 10^{-1}$ &( 5, 5) &\dec $-$3.8916016$\times 10^{-1}$  \\
 ( 4, 0) &\dec $-$1.2919922       &( 6, 1) &\dec $-$1.2828888$\times 10^{1}$ &( 4, 3) &\dec $-$1.0078125       &( 6, 5) &\dec $-$2.3588257        \\
 ( 5, 0) &\dec $-$3.4462891       &( 7, 1) &\dec $-$3.3050570$\times 10^{1}$ &( 5, 3) &\dec $-$2.7506104       &( 7, 5) &\dec $-$9.1123085        \\
 ( 6, 0) &\dec $-$7.1851196       &( 2, 2) &\dec $-$3.1250000$\times 10^{-1}$ &( 6, 3) &\dec $-$7.5901184       &( 6, 6) &\dec $-$5.0462341$\times 10^{-1}$  \\
 ( 7, 0) &\dec $-$1.6790197$\times 10^{1}$ &( 3, 2) &\dec $-$6.5625000$\times 10^{-1}$ &( 7, 3) &\dec $-$2.2107023$\times 10^{1}$ &( 7, 6) &\dec $-$3.6886940        \\
 ( 1, 1) &\dec $-$5.0000000$\times 10^{-1}$ &( 4, 2) &\dec $-$1.5449219       &( 4, 4) &\dec $-$3.1933594$\times 10^{-1}$ &( 7, 7) &\dec $-$6.8294907$\times 10^{-1}$  \\
\end{tabular}
\end{table}

\setdec 0.000000000000
\begin{table}
\squeezetable
\caption{Series coefficients for dimer expansions of the energy gap $E/J_{\perp}$ 
 of  singlet bound state, 
          triplet bound state, quintet antibound state,  and the lower and upper edge of 
          the continuum at $k=\pi$ for the the Heisenberg ladder.
Coefficients of $(J/J_{\perp})^n$
up to order $n=12$  are listed.}\label{tabserKpi}
\begin{tabular}{rrrrrr}
\multicolumn{1}{c}{$n$} &\multicolumn{1}{c}{singlet bound state} 
&\multicolumn{1}{c}{triplet bound state}     &\multicolumn{1}{c}{quintet antibound state} 
&\multicolumn{1}{c}{lower edge of continuum} &\multicolumn{1}{c}{upper edge of continuum} \\
\hline
  0 &\dec  2.000000000  &\dec  2.000000000  &\dec  2.000000000  &\dec  2.000000000  &\dec  2.000000000  \\
  1 &\dec $-$1.000000000  &\dec $-$0.500000000  &\dec  0.500000000  &\dec  0.000000000  &\dec  0.000000000  \\
  2 &\dec  0.750000000  &\dec  1.125000000  &\dec  1.125000000  &\dec  1.000000000  &\dec  2.000000000  \\
  3 &\dec  0.312500000  &\dec  0.312500000  &\dec  0.687500000  &\dec  0.250000000  &\dec  1.250000000  \\
  4 &\dec $-$0.203125000  &\dec $-$0.476562500  &\dec  0.148437500  &\dec $-$0.625000000  &\dec $-$0.500000000  \\
  5 &\dec $-$0.558593750  &\dec $-$0.742187500  &\dec $-$0.242187500  &\dec $-$1.031250000  &\dec $-$1.843750000  \\
  6 &\dec $-$0.356445313  &\dec $-$0.399414063  &\dec $-$0.198242188  &\dec $-$0.595703125  &\dec $-$1.119140625  \\
  7 &\dec  0.440856934  &\dec  0.444519043  &\dec  0.219665527  &\dec  0.648925781  &\dec  1.613769531  \\
  8 &                   &\dec  1.282394409  &\dec  0.294692993  &\dec  1.615997314  &\dec  3.436676025  \\
  9 &                   &\dec  0.964994431  &\dec $-$0.865842819  &\dec  1.012023926  &\dec  1.011138916  \\
 10 &                   &\dec $-$1.139695843  &\dec $-$3.052285552  &\dec $-$1.200890859  &\dec $-$4.719360987  \\
 11 &                   &\dec $-$3.099767812  &\dec $-$3.914894695  &\dec $-$2.788565993  &\dec $-$6.971628388  \\
 12 &                   &\dec $-$1.480682586  &\dec  0.070329791  &\dec $-$0.814231584  &\dec  0.478638977  \\
\end{tabular}
\end{table}

\begin{table}
\squeezetable
\setdec 0.0000000000000
\caption{The critical point (pole) and critical index (residue) obtained from
$[n/m]$ Dlog Pad\'e approximants to the series for the binding energy at $k=3\pi/5$
for the triplet bound state of the Heisenberg ladder. An asterisk denotes a defective
approximant.
}
 \label{tabLadderPade}
\begin{tabular}{rrrrrr}
\multicolumn{1}{c}{n} &\multicolumn{1}{c}{$[(n-2)/n]$}&\multicolumn{1}{c}{$[(n-1)/n]$}
&\multicolumn{1}{c}{$[n/n]$} &\multicolumn{1}{c}{$[(n+1)/n]$}&\multicolumn{1}{c}{$[(n+2)/n]$}
 \\
\multicolumn{1}{c}{} &\multicolumn{1}{c}{pole (residue)} &\multicolumn{1}{c}{pole (residue)}
&\multicolumn{1}{c}{pole (residue)} &\multicolumn{1}{c}{pole (residue)} &\multicolumn{1}{c}{pole (residue)} \\
\tableline
 n= 2 &                            &\dec 0.13342(2.100484) &\dec  0.13067(1.917138)$^*$ &\dec 0.13163(1.993828) &\dec 0.13173(2.002998) \\
 n= 3 & \dec 0.13169(1.999149)     &\dec 0.13172(2.001628) &\dec  0.13170(2.000083)     &\dec 0.13170(2.000146) &\dec 0.13170(1.999905) \\
 n= 4 & \dec 0.13171(2.000658)     &\dec 0.13170(2.000143) &\dec  0.13170(2.000097)     &\dec 0.13170(1.999987) &\dec 0.13170(1.999999) \\
 n= 5 & \dec 0.13170(1.999826)$^*$ &\dec 0.13170(1.999969) &\dec  0.13170(1.999997)   \\
 n= 6 & \dec 0.13170(2.000016)  \\
\end{tabular}
\end{table}


\begin{references}
\bibitem{jaanwork}
Jaan Oitmaa has developed some efficient computer programs to
generate a wide range of cluster for various lattices.
\end{references}
\end{document}